\documentclass{appolb}
\usepackage{epsf}
\usepackage[dvips]{color}

\begin{document}

\pagestyle{plain}

\title{Sign and amplitude representation of the forex networks}

\author{Sylwia Gworek$^1$, Jaros{\l}aw Kwapie\'n$^1$, Stanis{\l}aw Dro\.zd\.z$^{1,2}$
\address{$^1$ Institute of Nuclear Physics, Polish Academy of Sciences, Krak\'ow, Poland \\
$^2$ Faculty of Mathematics and Natural Sciences, University of Rzesz\'ow. Rzesz\'ow, Poland}}

\maketitle

\begin{abstract}

We decompose the exchange rates returns of 41 currencies (incl. gold) into their sign and amplitude components. Then we group together all exchange rates with a common base currency, construct Minimal Spanning Trees for each group independently, and analyze properties of these trees. We show that both the sign and the amplitude time series have similar correlation properties as far as the core network structure is concerned. There exist however interesting peripheral differences that may open a new perspective to view the Forex dynamics.


\end{abstract}

\section{Introduction}

In a series of our earlier papers~\cite{drozdz07,gorski08,kwapien09a,kwapien09b}, we showed that the currency exchange market has interesting properties if expressed as a network in different representations corresponding to different base currencies. Each representation is defined by a set of exchange rates sharing a common base which can form, via their cross-correlations, a binary or weighted network. Depending on a particular choice of the base, the network displays a hierachy with clusters of coupled currencies, with one or a few hubs and many peripheral nodes, among which a scale-free order can be observed. A node of the highest centrality is USD, which attracts a number of satellite currencies due to strong economical ties between the corresponding countries or explicite currency pegs. However, a careful inspection of the temporal evolution of the currency network structure, one observes a clear long-term trend according to which the USD node gradually loses its strength giving more freedom to its previously coupled neighbours. We showed that the only major currency which gains some importance owing to this trend is EUR.

In the present paper we take a closer look at some subtleties of the currency couplings. We study time series of daily exchange rates for a set of $N=41$ free-convertible currencies and gold~\cite{sauder}. Our data spans a decade-long interval from Jan 1, 1999 to Dec 31, 2008 ($T=2519$ trading days) and consists of all possible $N(N-1)=1640$ combinations of the echange rates B/X, where B is called the base currency (this rate expresses how many units of X one needs to buy 1 unit of B). The exchange rates $\Gamma_{\rm X}^{\rm B}(i), i=1,...,T$ obey the two fundamental relations:
\begin{eqnarray}
\Gamma_{\rm X}^{\rm Y}(i) = \Gamma_{\rm Y}^{\rm X}(i),\\
\Gamma_{\rm X}^{\rm Y}(i) = \Gamma_{\rm X}^{\rm Z}(i) \Gamma_{\rm Z}^{\rm Y}(i).
\end{eqnarray}
which reduce the effective dimensionality of the phase space. As usual in this type of analysis, due to strong nonstationarity of the exchange rates, we consider their logarithmic increments:
\begin{equation}
g_{\rm X}^{\rm B}(i) = \ln \Gamma_{\rm X}^{\rm B}(i+1) - \ln \Gamma_{\rm X}^{\rm B}(i).
\end{equation}
Each time series of $g_{\rm X}^{\rm B}(i)$ was preprocessed to remove possible artifacts; a low-pass filter at $10\sigma$ was also used in order to remove exceptionally high values which could influence outcomes of numerical analysis.

We want to avoid constructing of a single network from all the exchange rates since such a network would be large, information-overloaded, and difficult to comprehend (even though this kind of study can in principle be carried out - e.g.~\cite{mcdonald05}). Therefore we divide the whole set of time series into much smaller subsets consisting of the exchange rates with a given B. An advantage of this approach is that by fixing B and considering only the rates B/X we effectively eliminate this currency from an analysis. As a consequence, instead of the less convenient exchange rates, we can speak of the individual currencies and the relations between them.

For a given subset of time series, we construct an $N \times T$ data matrix ${\bf M}^{\rm B}$ and then calculate the correlation matrix ${\bf R}^{\rm B}$:
\begin{equation}
{\bf R}_{\rm XY}^{\rm B} = {1 \over T} {\bf M}_{\rm X}^{\rm B} \tilde{{\bf M}}_{\rm Y}^{\rm B},
\end{equation}
where $\tilde{\cdot}$ stands for matrix transpose. Matrix elements $R_{\rm XY}^{\rm B}$ are the correlation coefficients calculated for a pair of rates B/X and B/Y and can be considered a measure of couplings between currencies X and Y in the B-based representation of the market. By construction, $R_{\rm XY}^{\rm B} = R_{\rm YX}^{\rm B}$. The matrix ${\bf R}^{\rm B}$ completely defines the structure of an undirected network, in which currencies, expressed in terms of B, are nodes and the matrix elements are connection weights $\omega_{\rm XY}^{\rm B} = |R_{\rm XY}^{\rm B}|$. 

Such a network can be visualized as a complete graph, but this would not be an optimal graphical representation in our case of fully connected networks with 40 nodes and 780 internode links. Instead, we prefer to consider a Minimal Spanning Tree (MST)~\cite{mantegna99} - a subset of the whole network, consisting of all $N-1$ nodes and only $N-2$ links. In order to calculate MST, we change the correlation matrix into a distance matrix ${\bf D}^{\rm B}$ with elements defined by the metric:
\begin{equation}
d_{\rm XY}^{\rm B} = \sqrt {2 (1 - R_{\rm XY}^{\rm B})}.
\end{equation}
In general, $0 \le d_{\rm XY}^{\rm B} \le 2$; special cases are: $d_{\rm XY}^{\rm B}=0$ for identical signals, $d_{\rm XY}^{\rm B}=\sqrt{2}$ for uncorrelated signals (in the limit $T \to \infty$), and $d_{\rm XY}^{\rm B}=2$ for signals opposite in phase. Based on ${\bf D}^{\rm B}$ , MST is created by sorting its elements from the largest to the smallest, and connecting the nodes with links according to this ranking in such a manner that each pair of nodes may be connected only via a single path (no cycles allowed). This assures the dendric structure of the MST graph with a hierarchy of nodes such that the nodes which have in terms of $d_{\rm XY}^{\rm B}$ many close neighbours are higher in the hierarchy than the ones which are rather isolated. Although MST is only a specific choice of the network representation out of a broad spectrum of possible graphs, its usefulness in expressing the core information on the network structure has been shown in literature~\cite{bonanno04,mcdonald05,sieczka09,kwapien09a}. Using MST will make the graphical representation of our networks much more readable.

\section{Results}

\begin{figure}
\hspace{3.0cm}
\epsfxsize 6cm
\epsffile{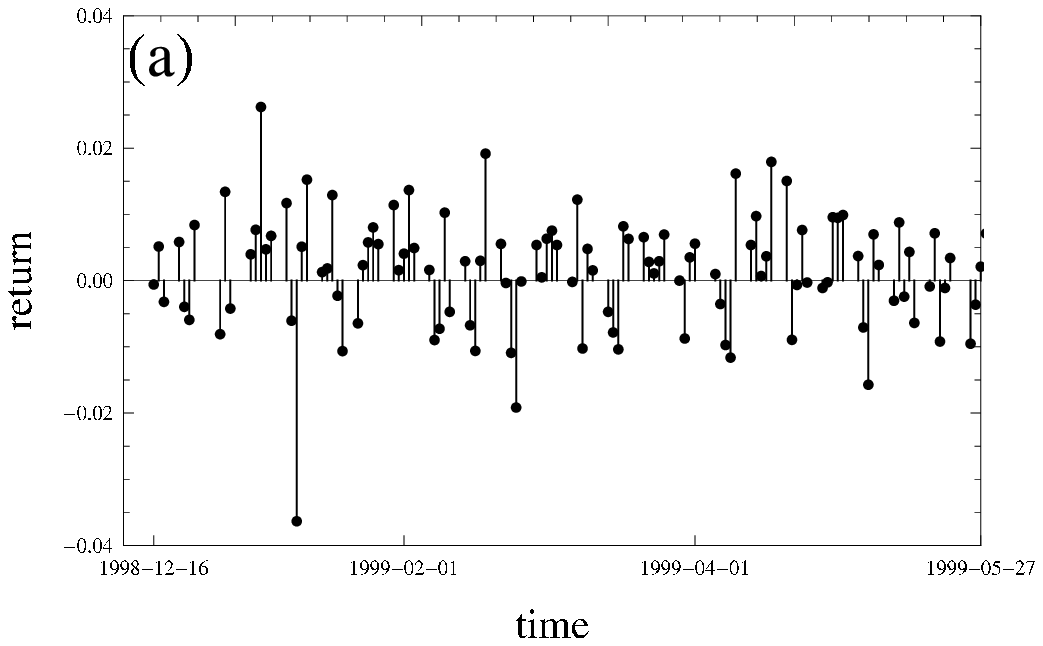}

\hspace{0.0cm}
\epsfxsize 6cm
\epsffile{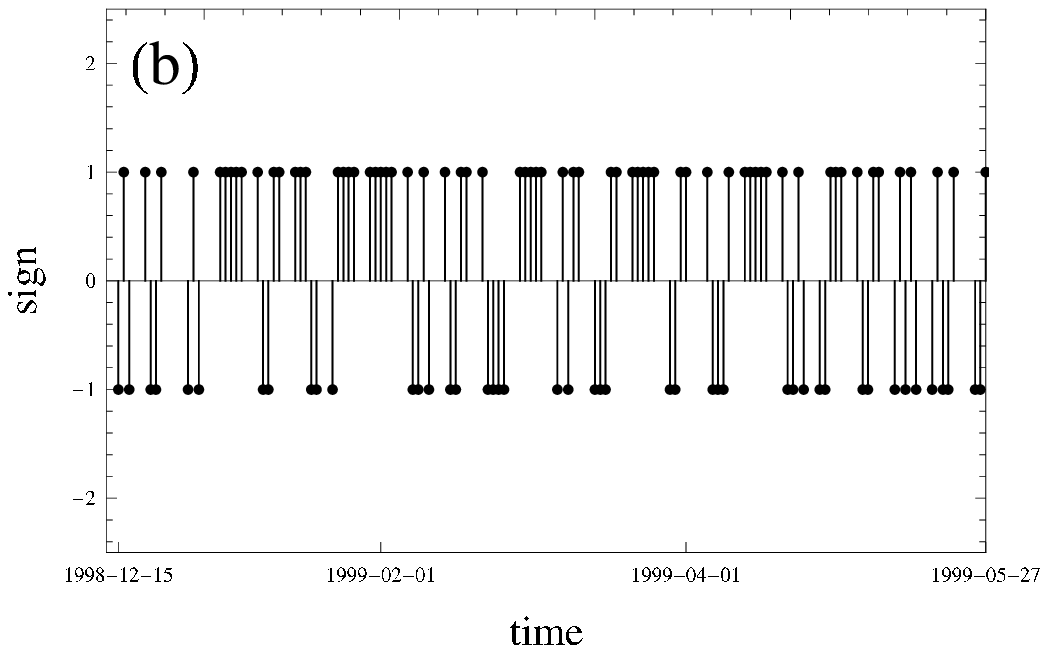}
\hspace{0.0cm}
\epsfxsize 6cm
\epsffile{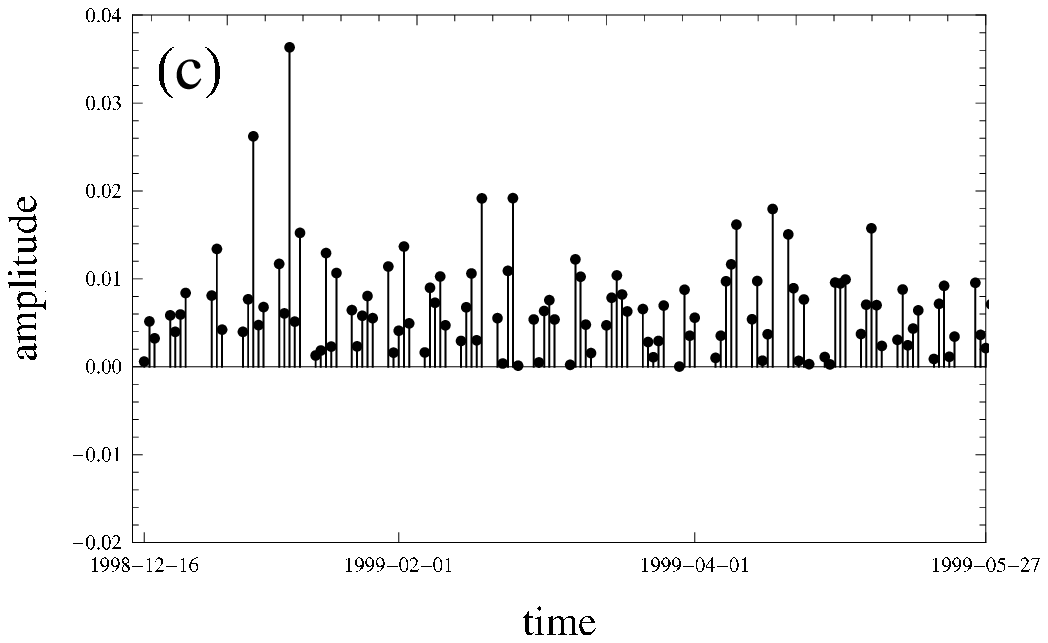}
\caption{Original time series of returns (a) decomposed into sign (b) and amplitude components (c). Gaps correspond to non-trading days.}
\end{figure}

In a preliminary inspection of the data we suppressed (to a threshold value of $\pm 10 \sigma$) a few exceptionally high fluctuations which can significantly affect the moments of a time series and, hence, also the correlation coefficients $R_{\rm XY}^{\rm B}$. Then we decomposed all the original time series  into their sign and amplitude components:
\begin{equation}
g_{\rm X}^{\rm B}(i) = {\rm sign} (g_{\rm X}^{\rm B}(i)) |g_{\rm X}^{\rm B}(i)| = s_{\rm X}^{\rm B}(i) a_{\rm X}^{\rm B}(i).
\end{equation}
In Fig.~1 we show the exemplary original signal and its components.
\begin{figure}
\hspace{2.0cm}
\epsfxsize 8cm
\epsffile{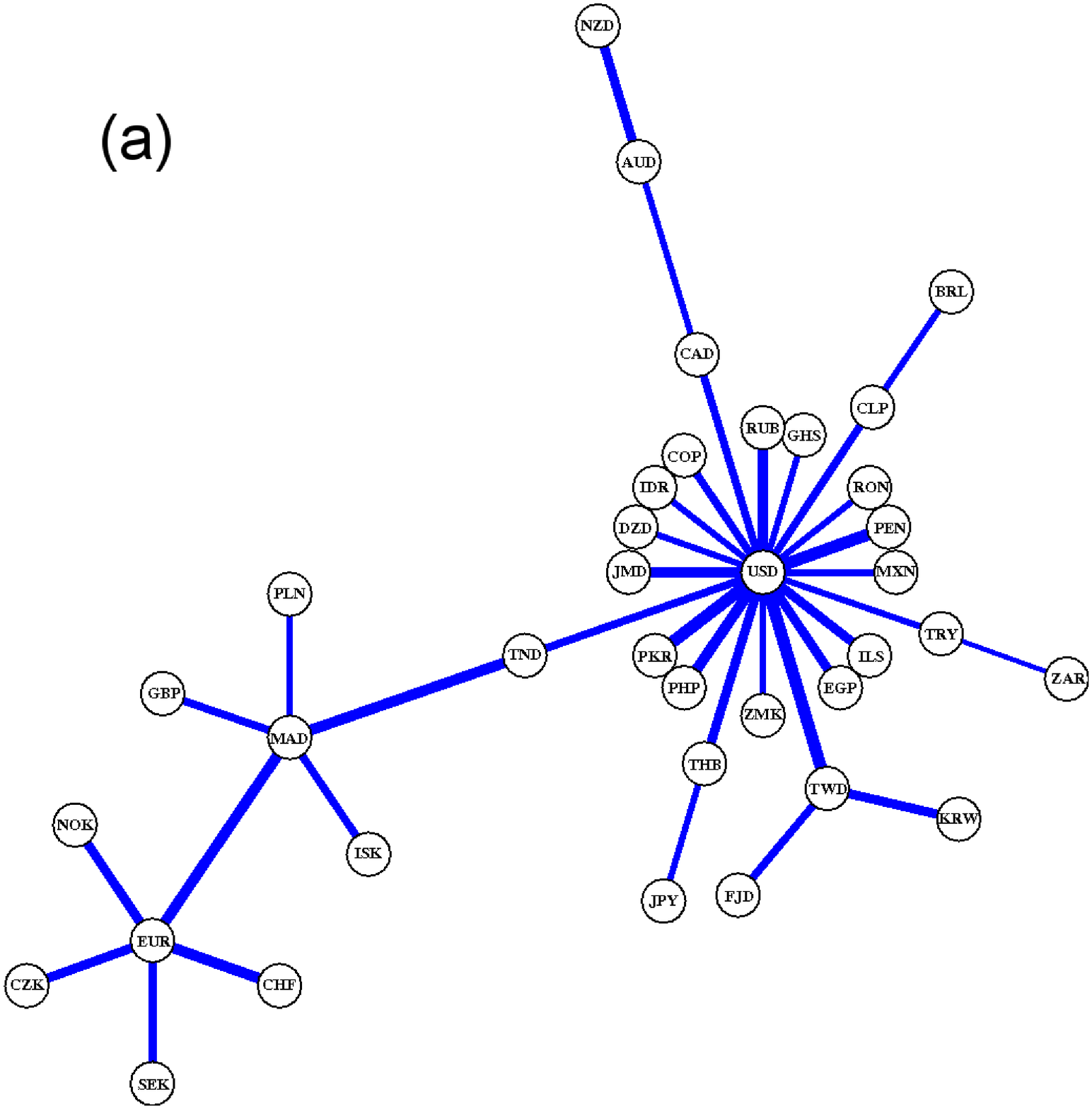}

\hspace{2.0cm}
\epsfxsize 8cm
\epsffile{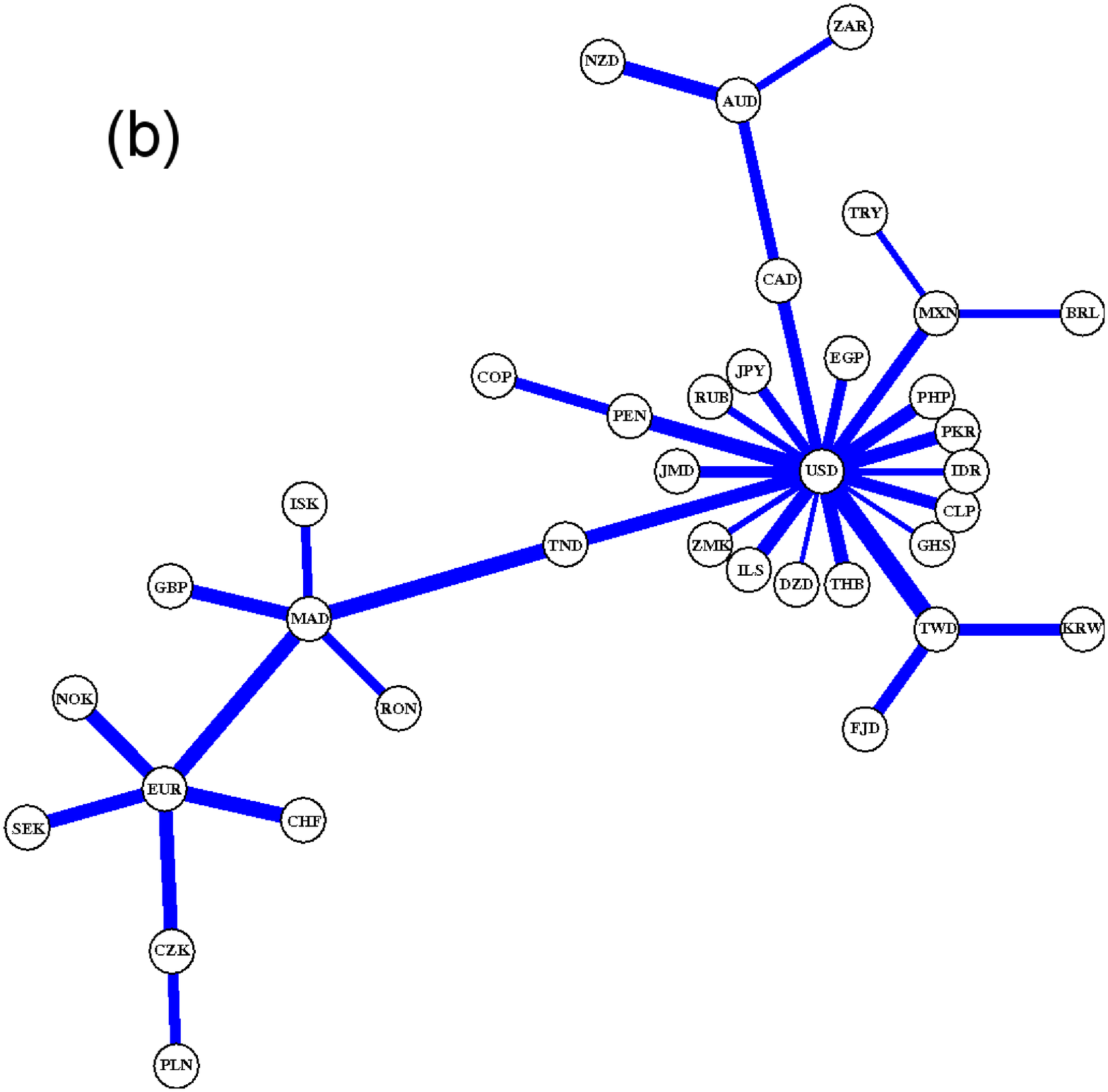}
\caption{Weighted MST graphs for the XAU-based network. The original signals of returns are decomposed into (a) signs (b) amplitudes. Width of internode connections is proportional to connection weights.}
\end{figure}
We start our presentation with the choice B$\equiv$XAU (gold). Gold, although at present cannot be considered a proper currency, is historically related to the currency market and its price evolution offers a convenient, independent reference frame for the whole market. MSTs for the XAU-based network are shown as weighted graphs in Fig.~2. Feature which is most striking is that for both types of signals all the internode connections have high weights. This is a manifestation of the fact that gold price is largely decoupled from the currency market, which in this representation acts as a large global cluster. At a more detailed level, the structure of internode connections among the nodes shows two prominent local clusters concentrated around USD and EUR, with the former clearly with a higher centrality than the latter. This picture remains qualitatively similar for the sign and the amplitude except for some minor restructuring of peripheral branches (which is a noise-like effect). This similarity together with the heavy weights suggest that gold price fluctuations have the same direction and magnitude if expressed in almost any currency. In other words, the gold price fluctuations are stronger (on average) than the fluctuations of all the exchange rates between currencies. This is an anticipated result.

\begin{figure}
\hspace{2.0cm}
\epsfxsize 8cm
\epsffile{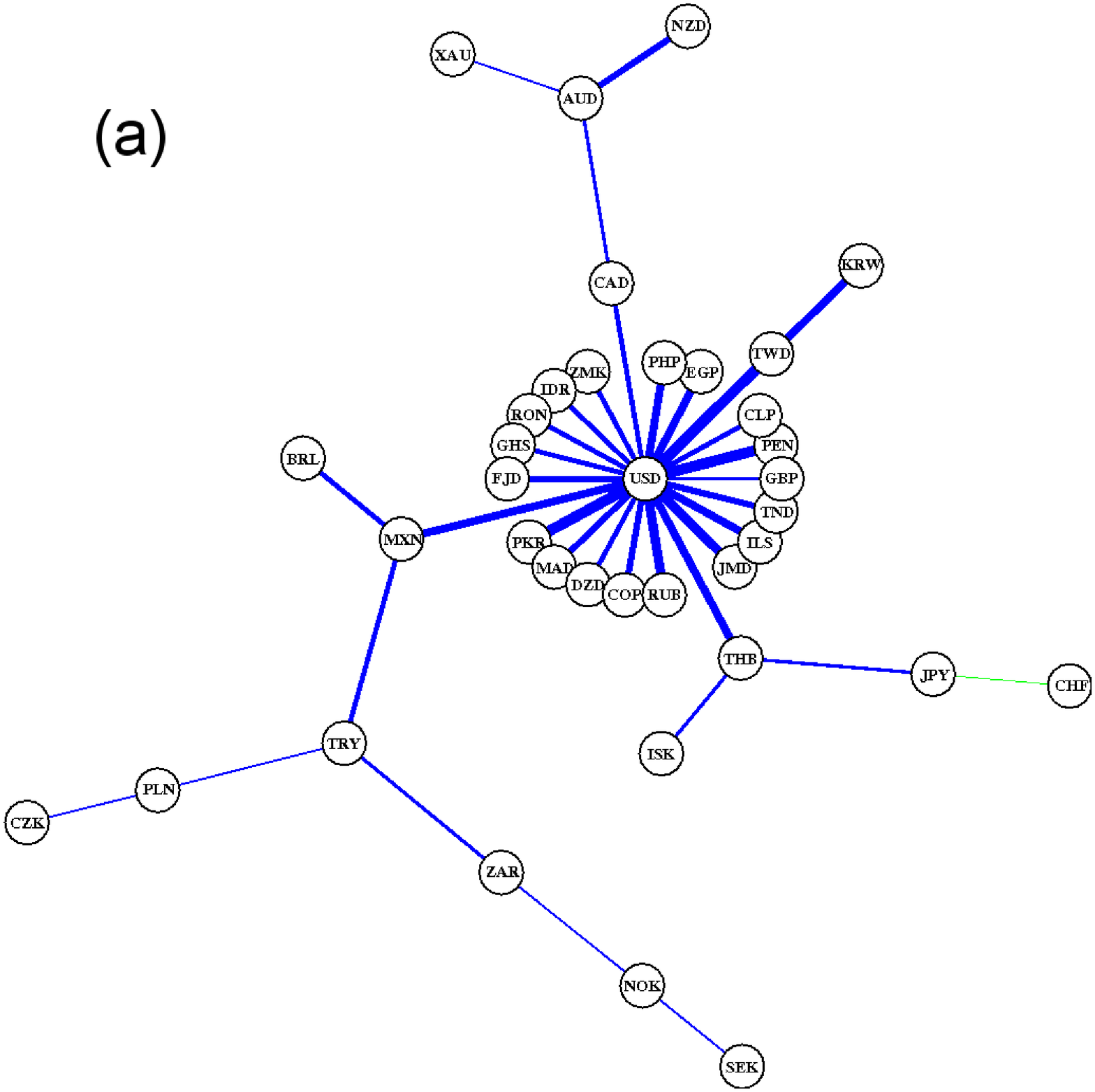}

\hspace{2.0cm}
\epsfxsize 8cm
\epsffile{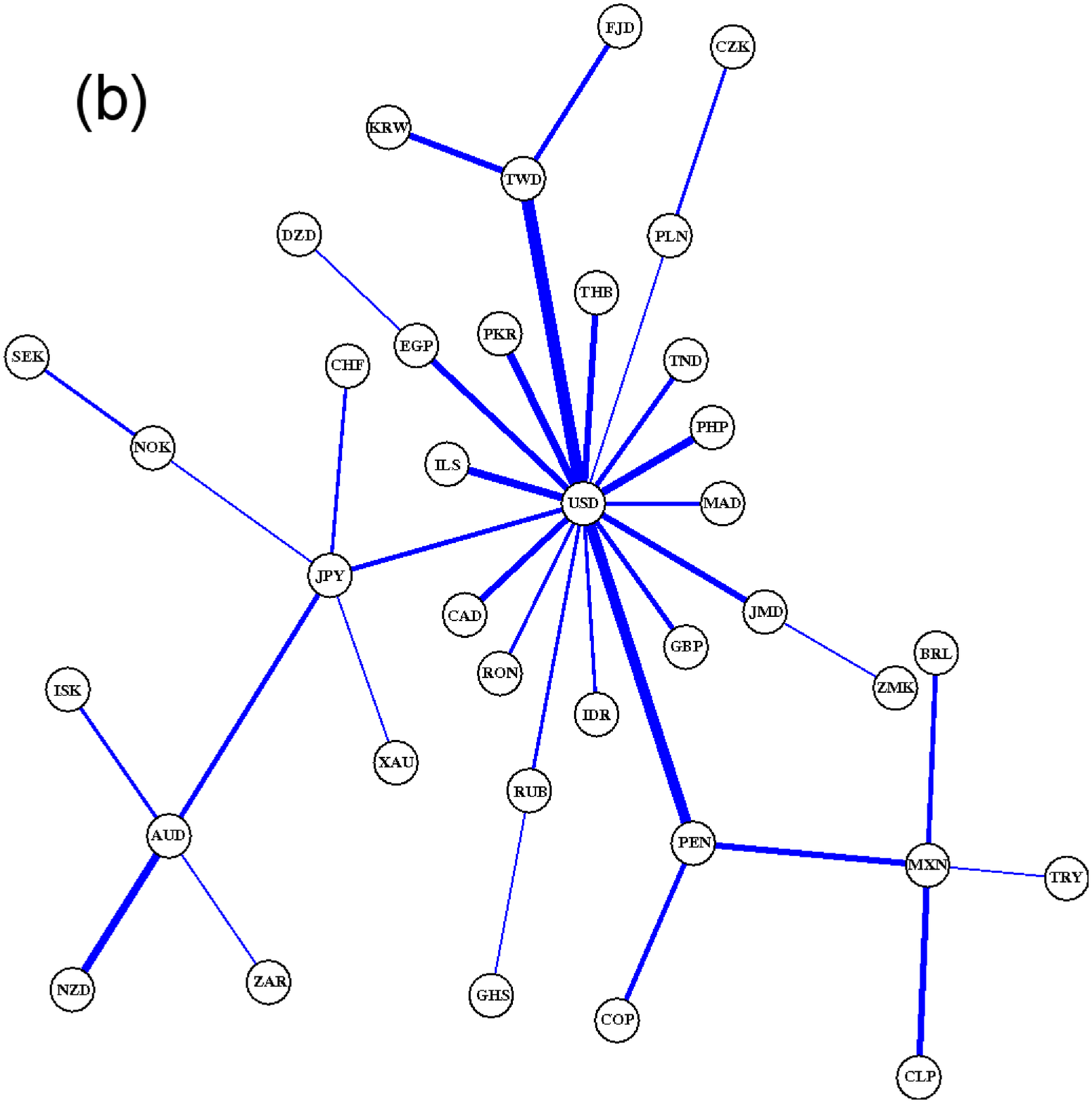}
\caption{Weighted MST graphs for the EUR-based network. (a) Signs. (b) Amplitudes.}
\end{figure}

The EUR-based networks for signs (a) and amplitudes (b) is presented in Fig.~3. We see that although average connection weights are much smaller than in the previous case of XAU-based network, the structure is to a large extent similar. There is also no prominent node reshuffling between the sign-based and the-amplitude based representations. USD occupies the central position in both MSTs with slightly more links in the case of signs (17 vs. 21). Average weights are comparable in both cases.

\begin{figure}
\hspace{2.0cm}
\epsfxsize 8cm
\epsffile{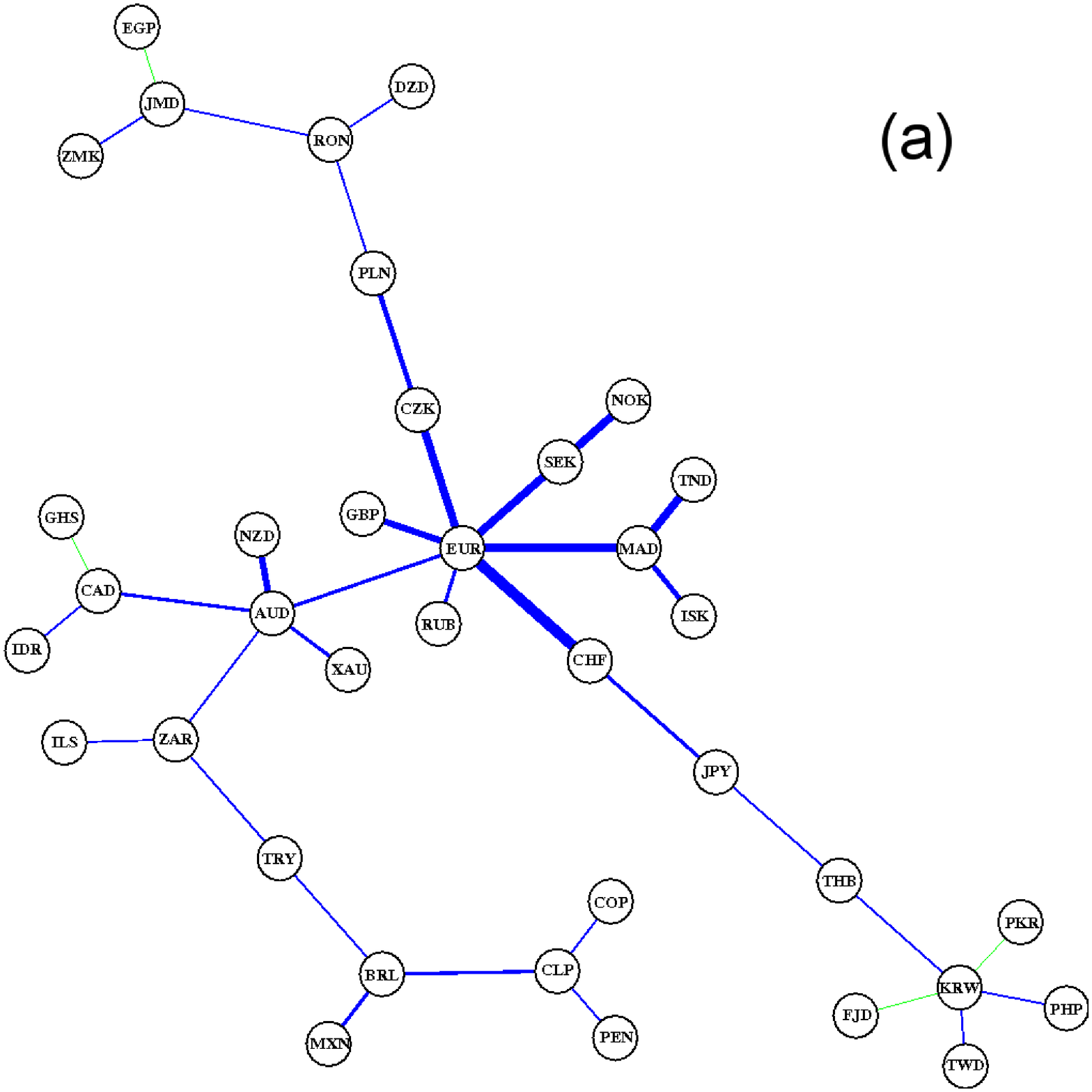}

\hspace{2.0cm}
\epsfxsize 8cm
\epsffile{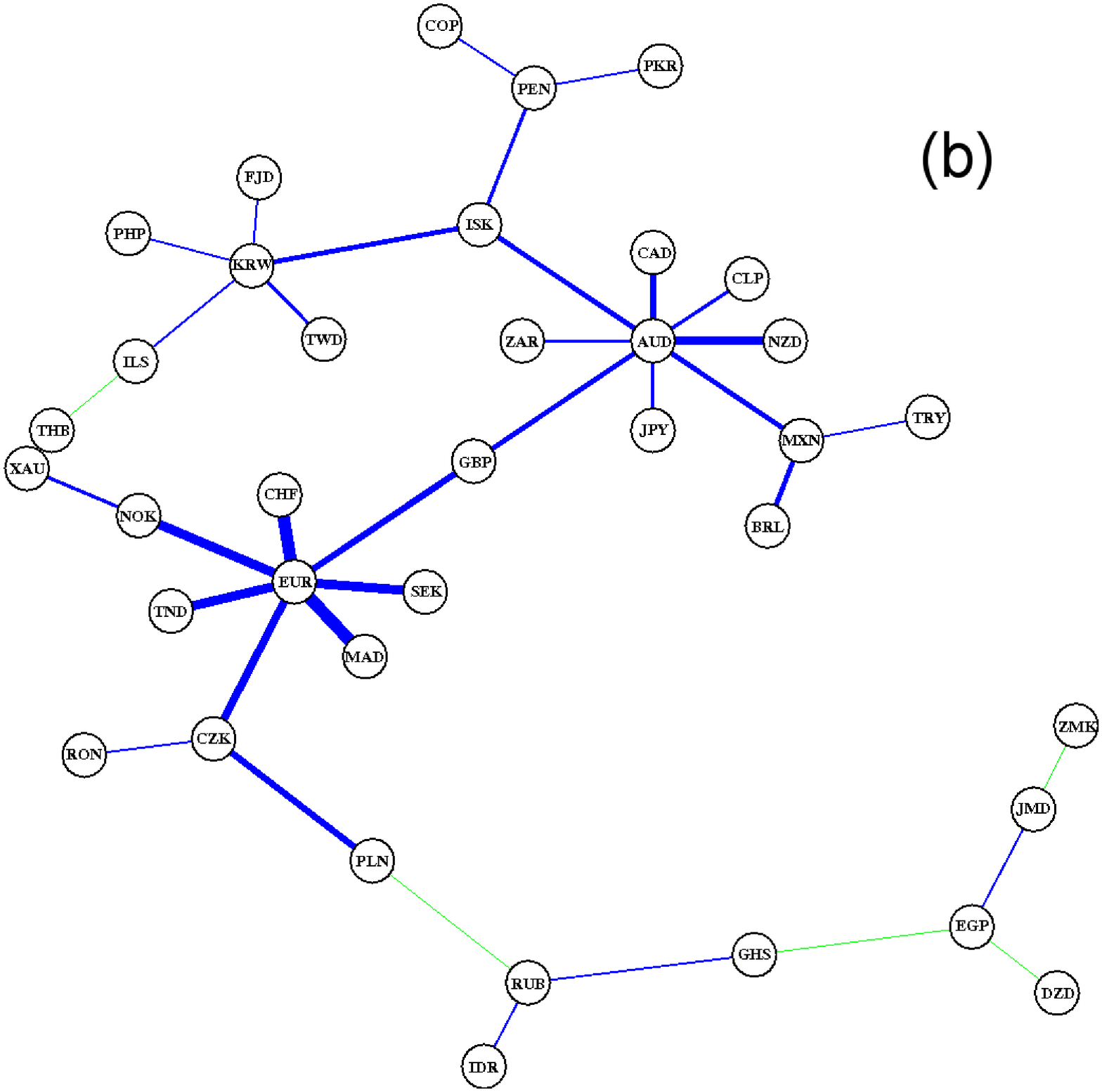}
\caption{Weighted MST graphs for the USD-based network. (a) Signs. (b) Amplitudes.}
\end{figure}

Let us now look at the MSTs as viewed from the USD perspective (B$\equiv$USD, Fig.~4). Such networks are expected to be least coupled and large group of nodes behave as if the network were random. Fig.~4 shows both the amplitude-based and the sign-based trees, which, as in Figs~2-3, do not present any meaningful differences between their structures. In both graphs EUR plays a role of the main hub, with the same multiplicity attracting all other European currencies as well as MAD and TND, which are in the EUR basin of attraction. The secondary hub, AUD, is also prominent with comparable number of links as EUR, but with smaller weights. For both types of signals, it attracts its standard neighbours: NZD, CAD, ZAR, and XAU. However, for the amplitude signals connections are stronger than for the sign ones. AUD also attracts more nodes in the case of amplitude signals among which notable examples are MXN and ISK which are not connected with this node if signed time series are considered (both signs and returns). In general, it is observed that for the USD-based network the amplitude signals do not develop so clear regional cluster structure as the signed signals.

As regards the secondary cluster structure, especially the South-East Asian and the Latin American cluster, it occurs that they are based on signs, at least to some extent, which is not the case for the strongest clusters of European and commodity-related currencies.

\begin{figure}
\hspace{-0.5cm}
\epsfxsize 6cm
\epsffile{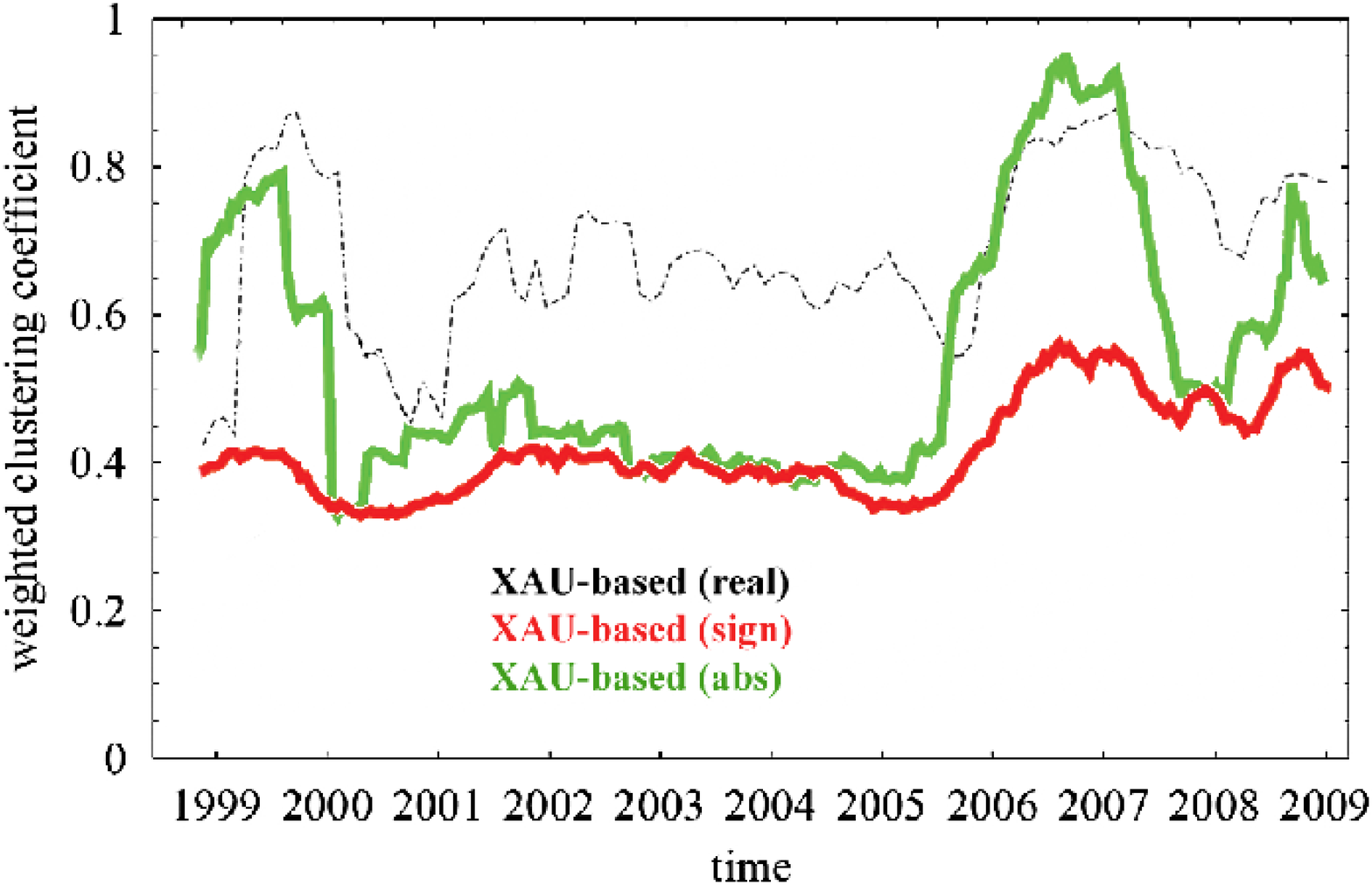}
\hspace{0.5cm}
\epsfxsize 6cm
\epsffile{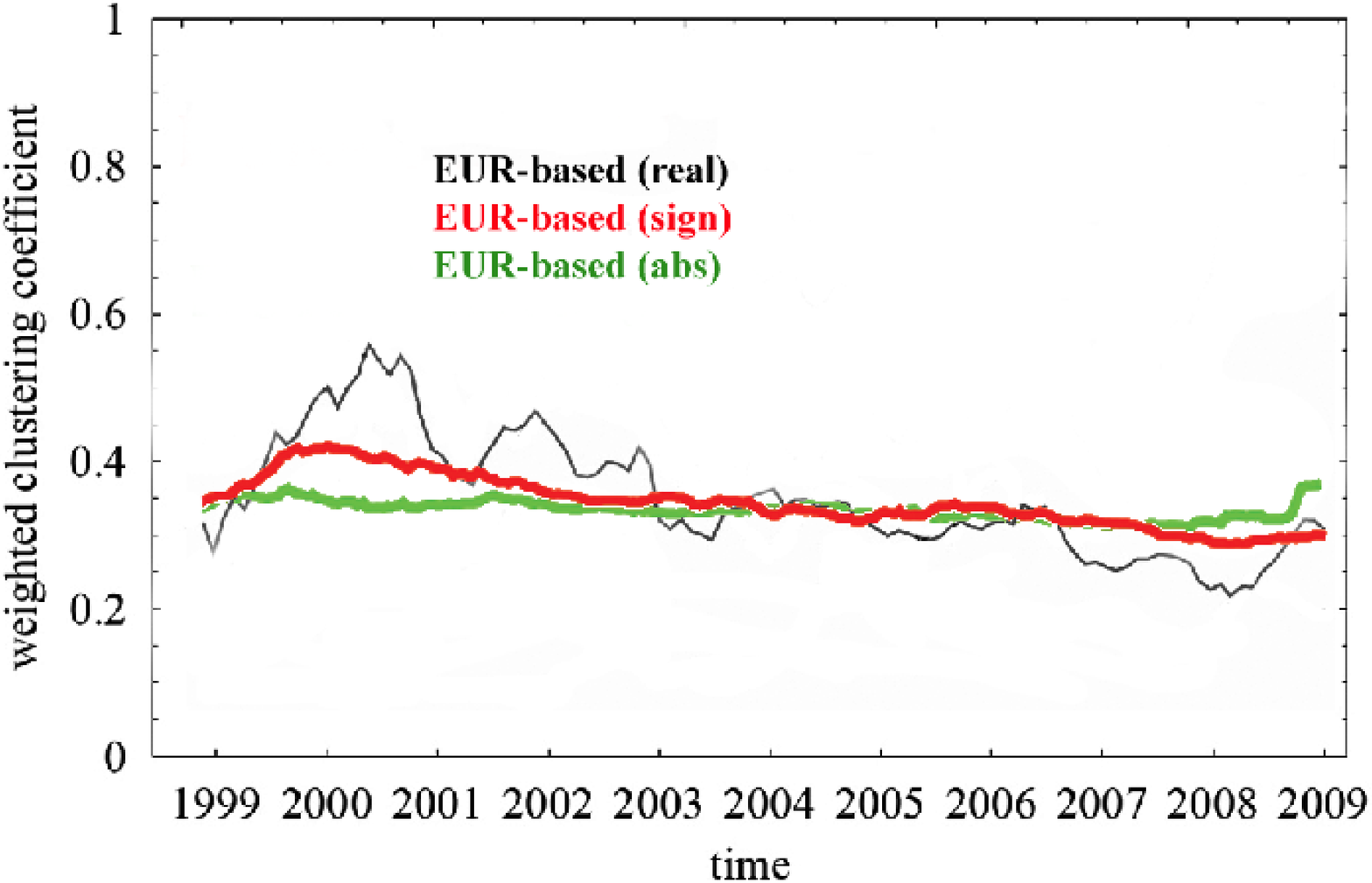}

\vspace{0.5cm}
\hspace{-0.5cm}
\epsfxsize 6cm
\epsffile{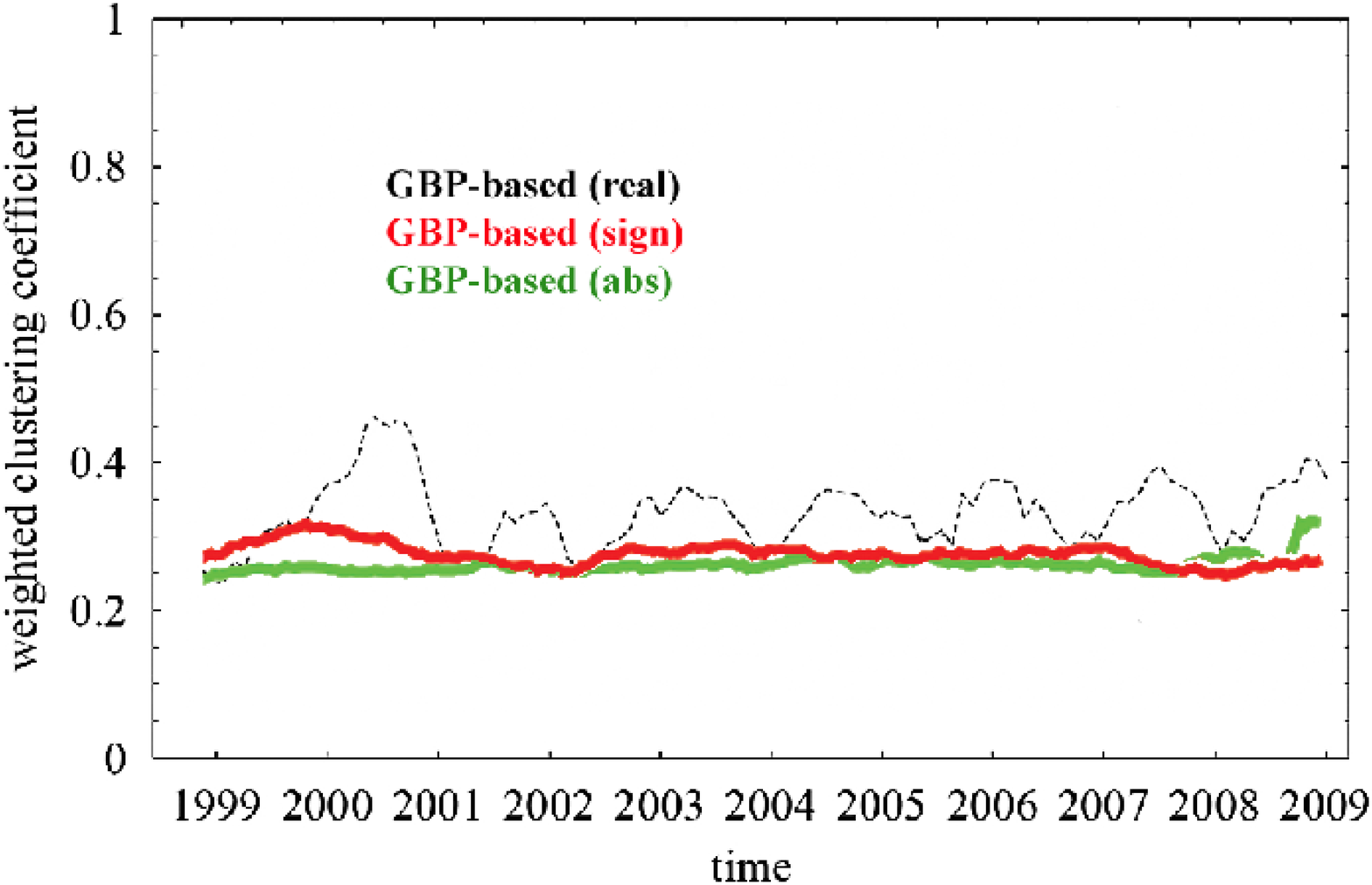}
\hspace{0.5cm}
\epsfxsize 6cm
\epsffile{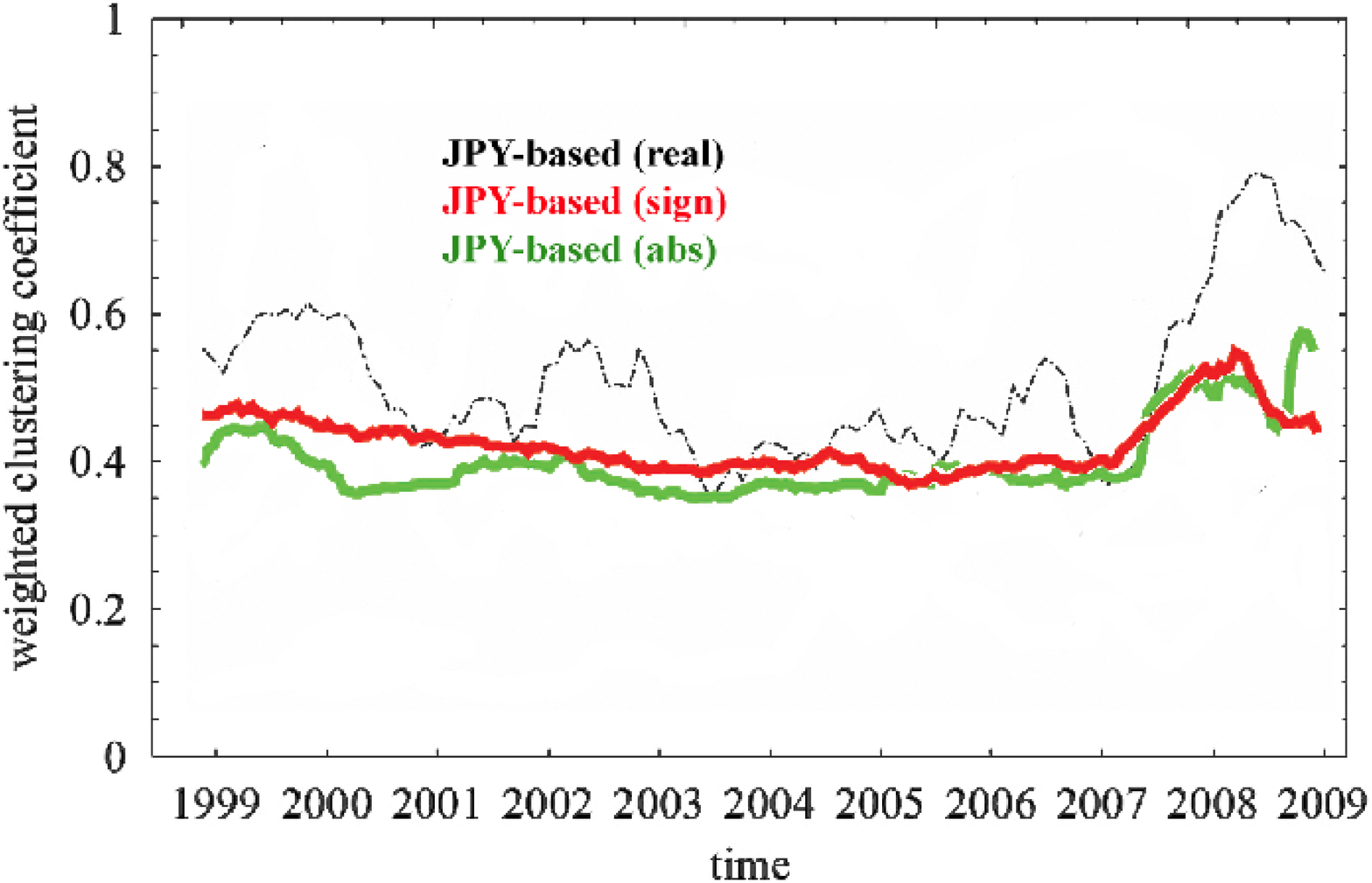}

\vspace{0.5cm}
\hspace{3.0cm}
\epsfxsize 6cm
\epsffile{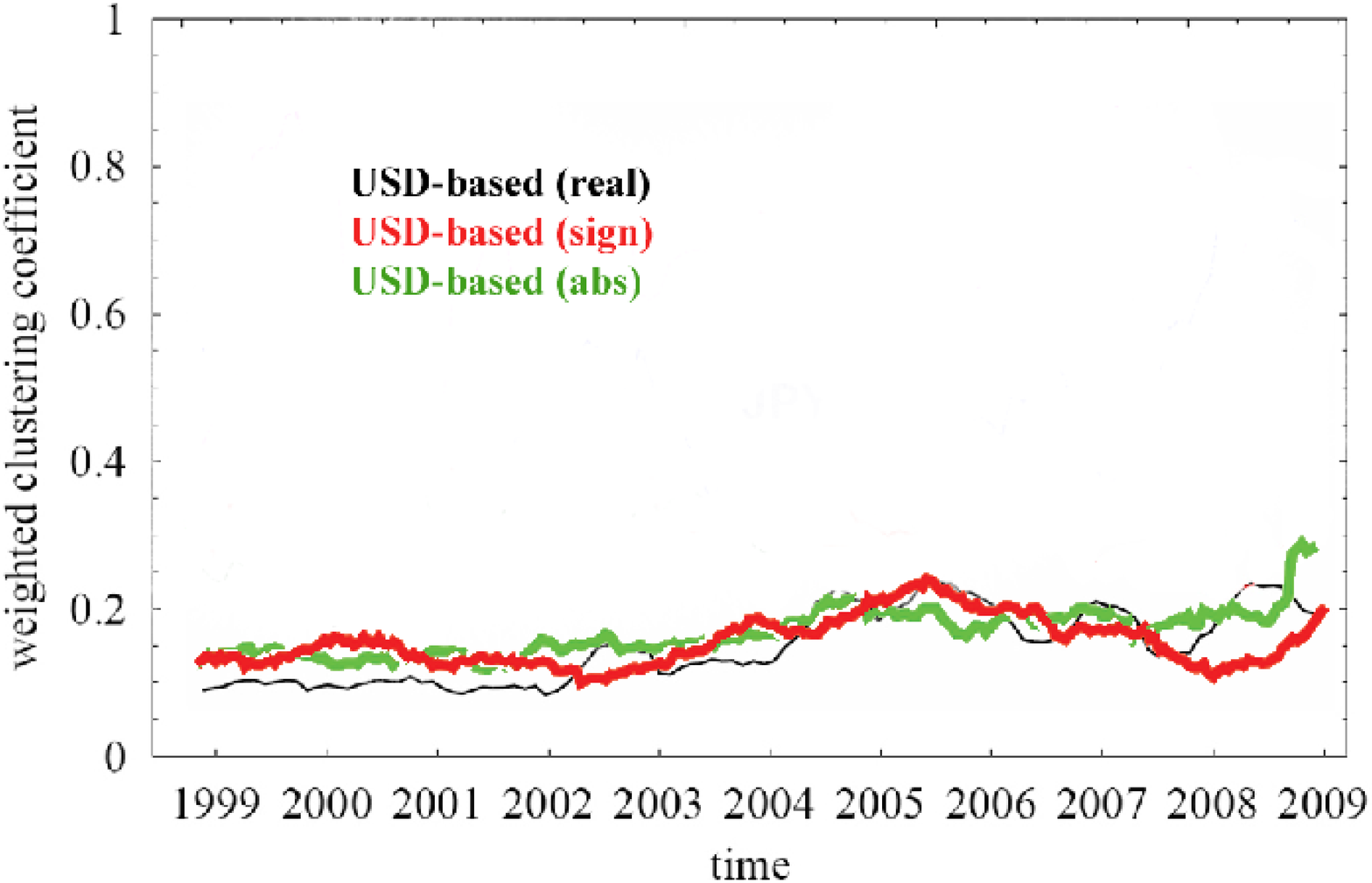}
\caption{Weighted clustering coefficient $\widetilde{C}^{\rm B}(t)$ for a few exemplary choices of the base currency. In each panel three types of data are presented: original signals (dashed black), signs (heavy red), and amplitudes (heavy green).}
\end{figure}

Let us now consider the MST topology in a quantitative manner by means of the characteristic path length $L^{\rm B}$ given by the formula:
\begin{equation}
L^{\rm B} = {1 \over N(N-1)} \sum_{{\rm X,Y : X \neq Y}} l^{\rm B}({\rm X,Y}),
\end{equation}
where $l^{\rm B}({\rm X,Y})$ stands for length of the path connecting nodes X and Y. Characteristic path length measures how compact is the tree in a given network representation B. Results are shown in Table 1 for a few base currencies. On average, the most distributed MSTs (largest $L^{\rm B}$) are formed by the signs, while the original signals (returns) and their absolute values are characterized by more compact MSTs. As expected, the longest paths characterize the USD-based trees which reflects the least correlated nature of this representation of the currency network. On the opposite end are the EUR-based trees. This is also understandable due to the fact that if EUR is chosen to be the base, there remains only one hub (USD) which makes the network highly centralized (Fig.~3). 

Another measure which can provide us with some insight into the network topology is the weighted clustering coefficient. It is defined for a complete network as~\cite{onnela05}
\begin{equation}
\widetilde{C}^{\rm B} = {1 \over N} \sum_{\rm X} \widetilde{c}^{\rm B}({\rm X}),
\end{equation}
where
\begin{equation}
\widetilde{c}^{\rm B} = {1 \over k^{\rm B}_{\rm X}(k^{\rm B}_{\rm X}-1)} \sum_{\rm Y,Z} \Big{(} \widetilde{\omega}^{\rm B}_{\rm X,Y} \ \widetilde{\omega}^{\rm B}_{\rm Y,Z} \ \widetilde{\omega}^{\rm B}_{\rm Z,X}\Big{)}^{\frac{1}{3}}, \ \ \ 
\widetilde{\omega}^{\rm B}_{\rm P,Q}= {\omega^{\rm B}_{\rm P,Q} \over {\rm max}_{\rm P,Q}[\omega^{\rm B}_{\rm P,Q}]}.
\end{equation}
In this case $K_{\rm X}^{\rm B}$ is the degree of node X.

\begin{table}[ht!]
\hspace{-0.4cm}
\begin{tabular}{|c|c|c|c|c|c|c|c|c|c|}\hline
Base: &CHF&CZK&EUR&GBP&GHS&JPY&PLN&USD&XAU\\
\hline\hline
$L^\mathrm{B}_{\tiny{\textit{\textrm{(return)}}}}$&1.63&1.73&1.53&2.33&1.55&1.95&1.99&4.10&1.65\\
$L^\mathrm{B}_{\tiny{\textit{\textrm{(sign)}}}}$&1.95&1.77&1.72&2.12&2.02&1.85&1.79&4.25&1.86\\
$L^\mathrm{B}_{\tiny{\textit{\textrm{(abs)}}}}$  &1.59&1.67&1.60&1.93&1.96&1.69&1.67&4.13&1.64\\
\hline\hline
$\widetilde{C}^\mathrm{B}_{\tiny{\textit{\textrm{(return)}}}}$&0.431&0.202&0.333&0.311&0.929&0.512&0.511&0.111&0.712\\
$\widetilde{C}^\mathrm{B}_{\tiny{\textit{\textrm{(sign)}}}}$&0.358&0.181&0.312&0.329&0.919&0.391&0.371&0.139&0.911\\
$\widetilde{C}^\mathrm{B}_{\tiny{\textit{\textrm{(abs)}}}}$&0.326&0.175&0.309&0.308&0.921&0.377&0.389&0.139&0.771\\
\hline\hline
\end{tabular}
\vspace{0.2cm}
\caption{The characteristic path length $L^\mathrm{B}$ and average weighted clustering coefficient $\widetilde{C}^\mathrm{B}$ for a few representative choices of the base currency B calculated for the full period 1998-2008 and deconstruct the real data into sign and amplitude components.}
\label{wspl}
\end{table}
q
Numerical values of $\widetilde{C}^{\rm B}$ are collected in Table 1 for the same base currencies as before. The results are highly B-dependent. For CHF, EUR and other continental European currencies, the clustering coefficient is highest for the original signals and lowest for the amplitudes, with the difference being largest for CHF and PLN. The same refers to JPY. In contrast, for the XAU-based and the GBP-based networks the highest clustering is observed for the signs, while both the amplitudes and the returns are significantly less clustered. USD is somehow peculiar in this respect, with equal level of clustering for the signs and the amplitudes, and a smaller level of clustering is observed for the returns.

\begin{figure}
\hspace{0.0cm}
\epsfxsize 6cm
\epsffile{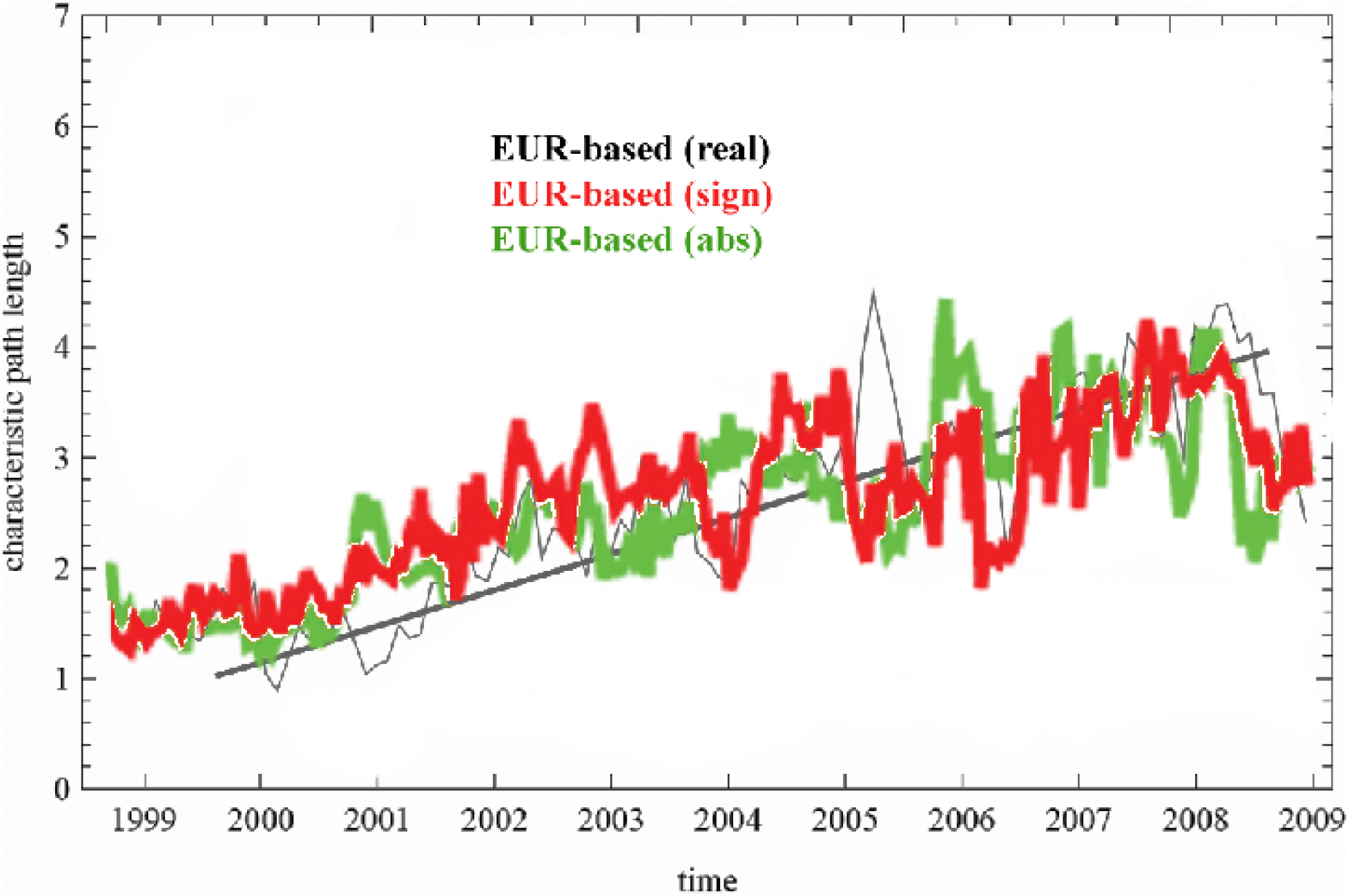}
\hspace{0.5cm}
\epsfxsize 6cm
\epsffile{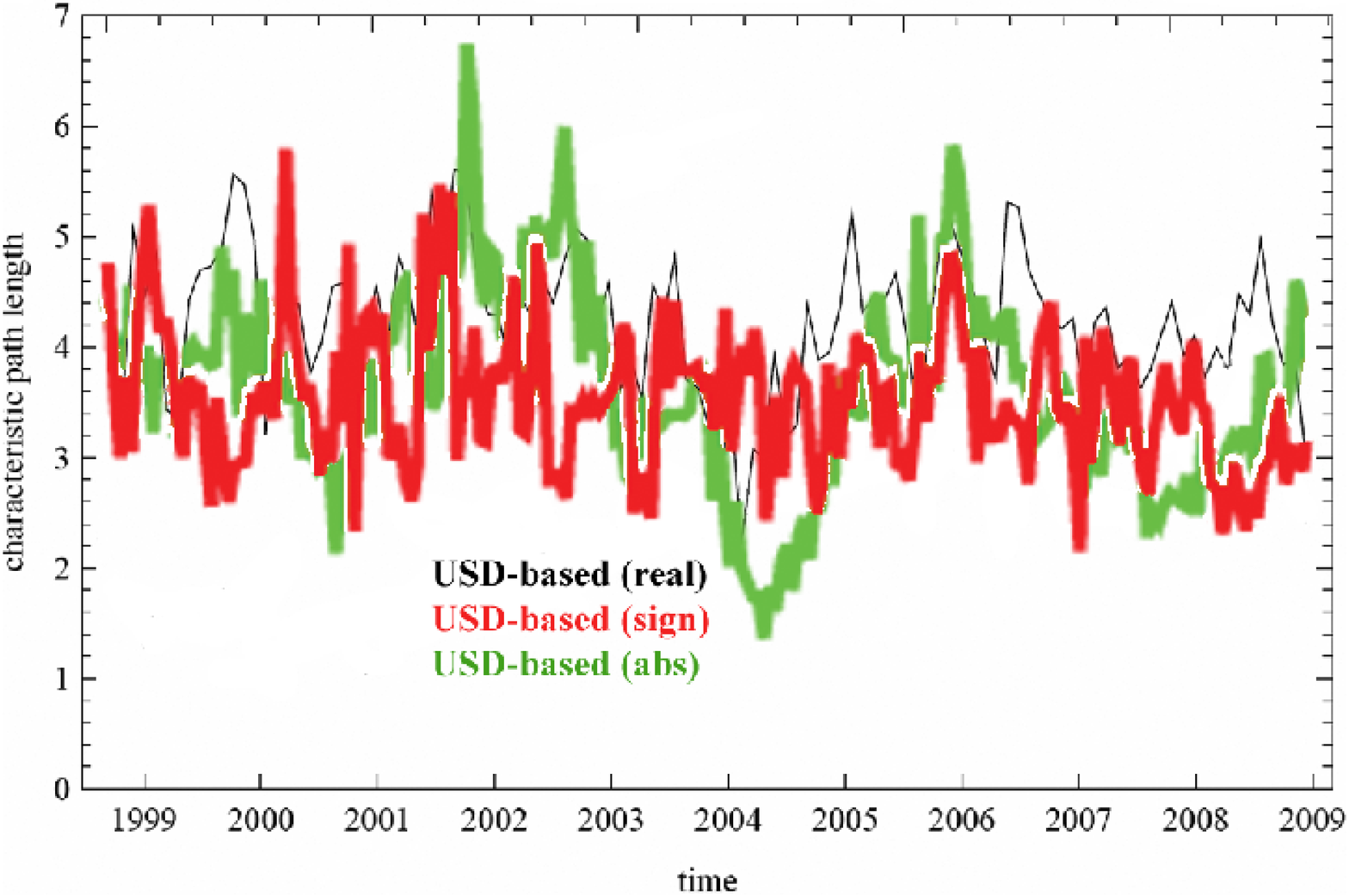}
\caption{Characteristic path length $L^{\rm B}(t)$ for the EUR-based and the USD-based network representations. Three types of data are presented: original signals (dashed black), signs (heavy red), and amplitudes (heavy green). Black solid line indicates a linear trend in $L^{\rm EUR}(t)$ for the returns.}
\end{figure}

We also consider the temporal stability of $\widetilde{C}^{\rm B}$ by using a moving window of 120 trading days (approx. 6 months). The results are presented in Figure 5 for 4 base currencies and XAU.
The most interesting observation is that for some choices of B, the network topolgy is much more stable in the case of the decomposed signals (the signs and the amplitudes) than in the case of original returns. This is true for GBP, JPY, and EUR, but not true for XAU and USD, These deconstructed signals exhibit sometimes different trends that the original complete signals, as it can be seen, e.g., in the case of the USD-based network. Nevertheless, in the case of the signs and the amplitudes, the EUR-based network exhibits a clear downward trend, similar to the one for the returns, indicating that the network viewed from the EUR perspective becomes more random-like (small clustering) than in the early years of EUR (Figure 6). This can be considered another manifestation of an increasing role of euro in the world trading system~\cite{drozdz07,kwapien09a}.

\section{Conclusions}

In summary, we studied the time series of currency exchange returns decomposed into their sign and amplitude components. We showed that the MST graphical structure does not change considerably for the components if compared with the original signals. However, more precise quantitative tools such as the characteristic path length and the weighted clustering coefficient indicate some restructuring of nodes and connection weights. We also observed that temporal stability of the networks for the components is better than for the returns. These are preliminary conclusions and  more work is needed in order to better understand their significance.

\end{document}